# Molecular structure in biomolecular condensates

Ivan Peran[1] and Tanja Mittag[1*]

[1]Department of Structural Biology, St. Jude Children's Research Hospital Memphis, TN, USA

[*]To whom correspondence should be addressed; email address: tanja.mittag@stjude.org

**Abstract**

Evidence accumulated over the past decade provides support for liquid-liquid phase separation as the mechanism underlying the formation of biomolecular condensates, which include not only "membraneless" organelles such as nucleoli and RNA granules, but additional assemblies involved in transcription, translation and signaling. Understanding the molecular mechanisms of condensate function requires knowledge of the structures of their constituents. Current knowledge suggests that structures formed via multivalent domain-motif interactions remain largely unchanged within condensates. Two different viewpoints exist regarding structures of disordered low-complexity domains within condensates; one argues that low-complexity domains remain largely disordered in condensates and their multivalency is encoded in short motifs called "stickers", while the other argues that the sequences form cross-$\beta$ structures resembling amyloid fibrils. We review these viewpoints and highlight outstanding questions that will inform structure-function relationships for biomolecular condensates.

**Introduction**

Liquid-liquid phase separation (LLPS) mediates many fundamental biological processes from cell signaling [1] to RNA metabolism [2], stress adaption [3] and transcription [4]. LLPS is rapidly becoming accepted as a key mechanism underlying the formation of biomolecular condensates [5,6], which include typical membraneless organelles such as nucleoli, nuclear speckles and stress granules but also less traditional compartments including heterochromatin [7,8], super-enhancers [4,9], the centrosomes [10], the pre- [11,12] and post-synaptic densities [13] and membrane receptor clusters [1,14,15]. Dysregulation of the (dis-)assembly of biomolecular condensates has been linked to cancer [16,17], neurodegenerative diseases [18,19] and aging [20]. While these observations suggest that phase separation is critically important for function, relatively few examples exist in which the functional requirement for phase separation has been firmly demonstrated. Contributing to this is a lack of understanding of molecular mechanisms mediated by LLPS and our current dearth in knowledge of the molecular structures within liquid dense phases.

LLPS is mediated by multivalent interactions of biomolecules [21-23] and characterized by a density transition and release of solvent [24] above the so-called saturation concentration. This gives rise to two coexisting phases, the dilute and the dense phase. Both phases have a system-specific concentration, which is independent of the total biomolecular concentration in the sample; an increase in the total concentration results in an increase in the volume fraction of the dense phase, at the expense of the volume fraction of the dilute phase, while both dense and dilute concentrations remain constant.

The multivalent interactions that mediate LLPS can be mediated by folded domains that are connected by disordered linkers, or they can be mediated by favorably interacting, so-called "sticky", residues or motifs within intrinsically disordered regions (IDRs) [25]. Both instantiations of multivalence can be conceptualized by the stickers-and-spacers framework, which has been used to describe the behavior of associative polymers [26] and has been adapted for proteins [27]. In this framework, stickers interact favorably with each other, whereas spacers neither interact with stickers nor with other spacers; their solvation properties, however, are important for determining whether crosslinking of the biomolecules is coupled to a density transition; multivalence without a density transition can lead to the formation of system-spanning networks of interacting molecules, i.e. gels, but does not result in the coexistence of dilute and dense phases [24]. The sticker valence, i.e. the number of interacting motifs within a protein, is anticorrelated with the saturation concentration, i.e. the higher the valence in a given system, the lower is its saturation concentration [21,27]. While many of these concepts directly connect with polymer theories and are useful to uncover general driving forces for phase separation, a complete understanding of the structural features of phase separation-mediating interactions and of the super-molecular structures within dense phases is required to fully appreciate the molecular function of biomolecular condensates. Here we review the current state of knowledge regarding biomolecular structures within condensates. We first review the structural properties of domain-motif interactions and then discuss the structures of intrinsically disordered regions within condensates while addressing the corresponding studies that have provided seemingly divergent yet important insights. We then touch upon the role of RNA structure in condensate formation and close with critical outstanding questions whose answers connect the structural and dynamical features of condensates with their emergent material properties and functions.

**Multivalent domain/motif interactions**

In many phase-separating proteins that have thus far been characterized, repeats of folded domains in one protein interact with repeats of linear motifs in a binding partner via multivalent interactions (Figure 1). Successive folded domains that are connected by linkers can be present in a single polypeptide chain, as in the classical examples first described by Rosen and colleagues [21]. For example, the protein Nck contains repeats of SH3 domains that interact with multiple proline-rich motifs (PRMs) in the protein N-WASP, and the protein PTB contains multiple RNA-binding domains and interacts with RNA, which is itself a multivalent molecule. The multivalence afforded by multiple folded interaction domains can also be obtained via oligomerization into discrete oligomeric species, as is the case in the nucleolar protein nucleophosmin (NPM1) [28] or by linear polymerization into higher-order structures, as in the case of the ubiquitin ligase adaptor Speckle-type POZ protein (SPOP) [16] and the RNA-binding protein TDP-43 [29]. In all of these examples, irrespective of the architecture that achieves multivalence, the proteins interact with binding partners that are themselves multivalent for binding to the interaction domains. These binding partners typically contain linear motifs and can be proteins or nucleic acids. However, the precise architecture determines the super-molecular structures that make up the dense phase and therefore determines its properties. SPOP oligomers bind multiple DAXX molecules, which results in brush-like structures which phase separate via DAXX-DAXX interactions [16,30]. This is in contrast to networks of SH3 domains that interact with PRMs, wherein single PRM chains can cross-link SH3 domains on different molecules directly (Figure 1).

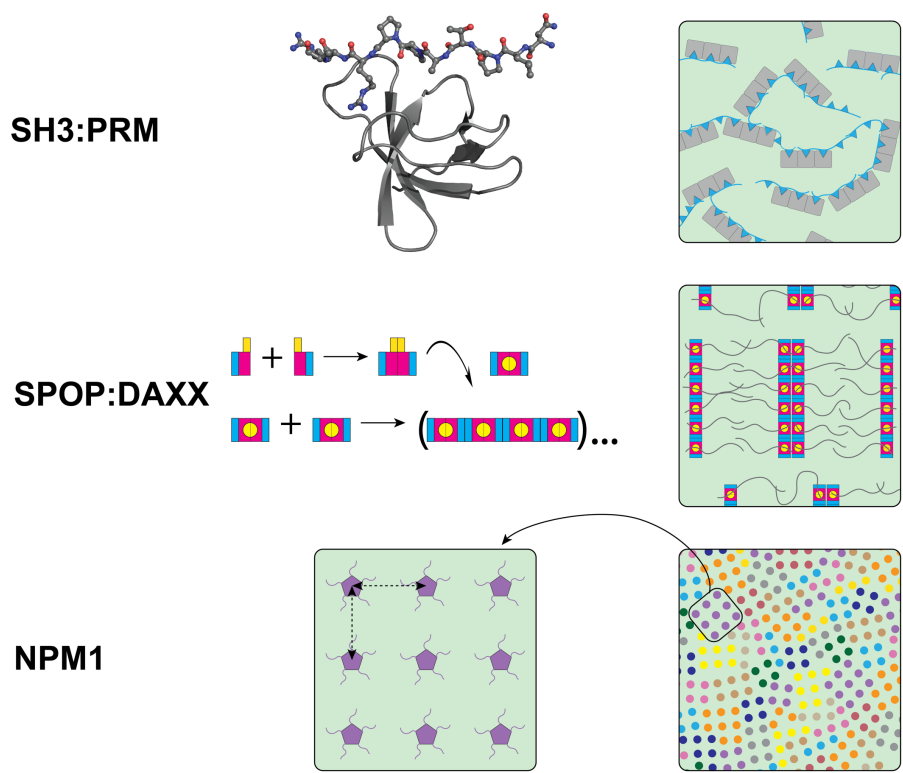

**Figure 1. Domain architecture determines super-molecular structure in dense phases. (Top row) Insight into dense phase structures via high-resolution structures of domain/motif interactions.** (Left) A high resolution crystal structure of an SH3 domain bound to a proline-rich motif (PRM) peptide. (PDB: 4WCI) [75]. (Right) A cartoon representation of a dense phase (green background) generated by multivalent SH3:PRM interactions. Three SH3 domains are connected by linkers in a single molecule (gray); a binding partner containing three proline-rich motifs (cyan) interacts with the SH3-protein via multivalent interactions. The two protein types crosslink each other. **(Middle row) Multivalency via linear polymerization.** (Left) SPOP monomers dimerize and the dimers polymerize into linear structures. The BACK dimerization domain is shown in cyan, the BTB dimerization domain in magenta, and the MATH substrate binding domain in yellow. The curved arrow indicates a 90° rotation of a SPOP dimer onto its side. (Right) SPOP oligomers bind several DAXX molecules via their MATH domains, giving rise to "brush-like" structures. These brushes are crosslinked via DAXX interactions [16,30]. **(Bottom row) Local order in a liquid dense phase.** The distances between NPM1 pentamers (magenta) repeat (left), giving rise to locally ordered arrays of Npm1 molecules within the dense phase. Npm1 molecules are crosslinked by arginine-rich peptides (not shown). (Right) Over greater distances, the anisotropic arrangement of arrays (each shown in different colors) gives rise to global disorder and liquid-like behavior [28].

The domain-motif interactions in these systems are modular in nature and typically well understood, given that high-resolution structures of individual complexes are often available (Figure 1). The identical structures, strung together by disordered linkers, are also assumed to give rise to the higher-order complexes formed in the multivalent case. It is these higher-order complexes that undergo phase separation due to their reduced solubility compared to individual protein molecules and small complexes. The remaining dilute phase also contains higher-order complexes that coexist with the dense phase [21], and their sizes depend on the effective affinities between the binding partners and the saturation concentration of the system.

The assumption that the structures of building blocks are identical in the dilute and dense phases may be incorrect due to steric constraints imposed by their multivalence, or if the stability of the domains/complexes or their structures are affected by the altered solvation properties of the dense phase. Further, the stability and structure may be influenced by differential partitioning of ions and biomolecules between the light and dense phases. In fact, single molecule FRET experiments demonstrated the presence of distinct NPM1 conformations populating the light versus dense phases. In dilute solution, the A2 tract, a disordered region in NPM1, forms a compact conformation stabilized by electrostatic interactions; in the dense phase the A2 tract is expanded as a consequence of interactions with positively charged nucleolar proteins [28].

Information pertaining to the super-molecular structure of these complexes within dense phases is sparse such that the relative orientations and distances of individual domain/motif complexes in dense phases are largely unknown. The only available information stems from SANS data of the dense phase formed by a truncated form of NPM1 and an arginine-rich, nucleolar-derived peptide [28]. The SANS data shows diffraction peaks, thus implying regularly repeating distances in the dense phase, which can be interpreted as distances between neighboring domains or NPM1 oligomers. This implies a regularly repeating pattern of the NPM1 molecules and at least local order in the dense phase, even if the dense phases are disordered on longer length scales (Figure 1).

It is intuitive to expect that the super-molecular structures within dense phases are dependent on system-specific properties such as linker lengths and sequences. A full description of the structures in the dense phase, the lifetime of individual interactions, and how the dynamics of one building block is correlated with that of its neighbors would open the door to a molecular understanding of the emergent material properties of dense phases. We expect that an integrative approach utilizing solution- [28,31] and solid-state NMR spectroscopy, scattering and single molecule fluorescence techniques [32] together with molecular simulations [33,34] will facilitate our understanding of how behavior at the molecular level correlates with the emergent material properties of condensates. Cryo-EM tomography [3] may play a particularly important role as it will allow bridging between structural biology of condensates in vitro and in cells.

**Low complexity domains mediate phase separation**
An analysis of the components of membraneless organelles has shown an enrichment of proteins containing intrinsically disordered low complexity domains (LCDs) [21,35,36]. The composition of these sequences is biased and typically enriched with few select amino acids and the chains fail to fold into well-ordered three-dimensional structures. LCDs mediate the phase separation of many proteins and are often necessary and sufficient for phase separation [37-39]. Given these observations, it is likely that LCDs play important roles in the formation of biomolecular condensates and tune their material properties. Among the LCDs used as models for phase separation are Fused in Sarcoma (FUS) and the RNA-binding proteins hnRNPA1, hnRNPA2 and TDP-43. Their LCDs are termed prion-like domains or PrLD due to their compositional similarity to yeast prion proteins; they are enriched in asparagine, glutamine, tyrosine and glycine residues [40]. Two views regarding structures of LCDs within condensates have emerged. One side argues that regions within the LCD fold into a β-sheet structure and that many such structures assemble into amyloid-like fibrils with a cross-β architecture. The opposing view maintains that LCDs remain largely disordered within the dense liquid phase (Figure 2).

**Cross-β structure of LCDs in fibrils and hydrogels**
A number of reports have shown that LCDs from RNA-binding proteins and similar sequences can generally form hydrogels, which are composed of amyloid-like fibrils [21,41-44]. Characterization of fibrils showed that, despite the presence of cross-β architecture, these structures are distinct from conventional amyloid fibrils such as those of α-synuclein and Aβ [45-50] (Figure 2). Fibril formation of LCDs is often reversible and fibrils dissolve upon dilution, at high temperature or upon treatment with detergents [21,41,51]. Compared to α-synuclein and Aβ, the sequences are enriched in polar amino acids while lacking hydrophobic residues. A combination of solid- and liquid-state NMR spectroscopy were used to elucidate the structure of FUS fibrils; they consist of a β-sheet containing core of 57 residues (23% of the total sequence) within the N-terminal half, while the remainder of the chain (77%) remains disordered [41]. The core region lacks a hydrophobic interface and contains residues that can hydrogen bond to one another.

Recently, short motifs were identified within the LCDs of RNA-binding proteins that can mediate fibrillization and they were termed LARKS (low-complexity aromatic-rich kinked segments) [42,52]. Powder diffraction showed that when isolated as peptides, LARKS formed cross-β structure that was characterized by x-ray crystallography and micro-Electron Diffraction at atomic resolution. The structures are stabilized by aromatic stacking and hydrogen bonds. Kinks within the structures not only allow for close approach of backbones and favorable hydrogen bonding and van der Waals interactions between sheets, but also prevent the sidechains from interdigitating across the β-sheet interface so that less surface area is buried when compared to conventional amyloid fibrils. These structures may thus explain the reversibility of fibrillization and their properties have been rationalized to resemble the transient interactions and high mobility of protein molecules in RNA granules revealed by fast fluorescence recovery after photobleaching (FRAP) [37]. An analysis of the human proteome suggested that LARKS are enriched in hundreds of proteins in biomolecular condensates, the nuclear pore complex and in the extracellular matrix [42], suggesting that the LARK fibril structures may represent relevant modes of interactions in these mesoscale structures in cells. In FUS, LARKS are distributed throughout the sequence and not only in the β-sheet containing core of the fibrils characterized by solid-state NMR. [41,42].

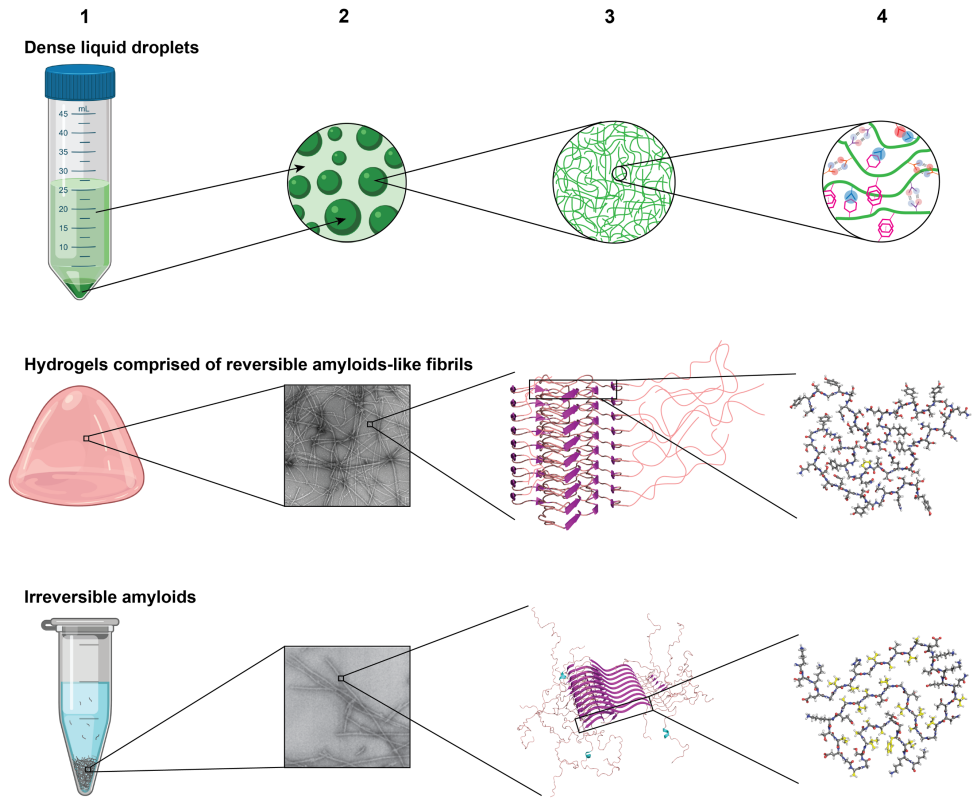

**Figure 2. LCDs can form different types of assemblies. (a)** LLPS of intrinsically disordered LCDs. From left to right: (1) A sample with a large dense drop at the bottom of the tube (dark green) overlaid by the dilute phase (light green) is prepared from (2) many small micron-sized droplets (dark green) that fuse. (3) The protein chains remain largely disordered in the dense phase (green mesh). Interactions driving phase separation include hydrophobic, π-π interactions (e.g. between aromatic residues shown in magenta stick representation), cation-π interactions between aromatics (magenta sticks) and positively charged residues (blue sticks), polar interactions (violet sticks), electrostatic interactions between positive (blue sticks) and negative (red sticks) charges, and hydrogen bonding (orange sticks). **(b)** Hydrogels of LCDs. From left to right: (1) Many low complexity domains such as that of FUS can assemble into hydrogels. (2) The FUS hydrogel is composed of amyloid-like fibrils. (3) Solid-state NMR structure of the core region of FUS that assembles into long filaments with a cross-β architecture. The remaining 77% of the LCD sequence remains disordered, indicated as red lines. (4) The atomic structure of a FUS monomer within a fibril reveals short in-register β-sheets separated by loops. Of note is the absence of all but one hydrophobic residue (yellow). **(c)** Irreversible amyloid fibrils formed by typical amyloidogenic proteins, not LCDs. (1) Cartoon depiction of irreversible fibrils sedimented in an Eppendorf tube. (2) TEM of α-synuclein fibrils. (3) Structure of the components of the fibrils indicates a long, continuous β-sheet core flanked by disordered regions. (4) Atomic structure of the core of an α-synuclein monomer within a fibril shows an in-register β-sheet stabilized by hydrophobic contacts involving Val, Ile, Ala and Phe residues (yellow). (Some illustrations were made in BioRender (biorender.com).)

Complementary work has identified additional motifs within the LCD of hnRNPA1 and FUS that form reversible fibrils with cross-β structure; these motifs were termed hnRACS (reversible amyloid cores) [43,44].The core of the fibril consists of Gln residues that form intersheet hydrogen bonds. An FG motif on one side of the core creates a kink that allows the aromatic residue to reach out and engage in a network of interactions, which is required for gelling, i.e. for the branching of fibrils and their arrangement as three-dimensional networks. The reversibility and instability of the fibrils seem to result from stacking of negatively charged Asp residues along the fibril axis; in accordance with this, replacement of Asp with Val abolishes the reversibility of fibrillization and may explain common disease mutations [44,53]. However, the authors observed both reversible and irreversible fibrils that are formed from the same sequence. Future studies should address how reversible fibrils that lack a hydrophobic interface convert over time into irreversible fibrils [44].

A series of reports has presented compelling evidence that features required to form fibril structures are also functionally important inside cells [41,54,55]. Many LCDs of RNA-binding proteins that are associated with stress granules have affinities for FUS hydrogels [21]. Importantly, sequence features that mediate the ability to form fibrils/hydrogels and mediate binding of monomers to hydrogels are also functionally important. This includes the tyrosine residues punctuating the FUS LCD sequence and the methionine residues punctuating the LCD of Pbp1, the yeast ortholog of Ataxin-2. The tyrosine residues in FUS are not only required for hydrogel formation but for transcriptional activity [55] and the methionine residues for redox sensitivity of Pbp1 condensates [54]. This represents a classical biochemical validation of these contacts. Based on these observations, it was postulated that the fibril-containing hydrogel bears similarities to cellular RNA granules. Phosphorylation of residues within the β-sheet core of FUS significantly disrupt hydrogel binding and causes droplet dissolution, further arguing for the importance of the structured core for incorporation into hydrogels and droplets [41]. Further, Li, Liu and colleagues demonstrated that removal of hnRACs from the LCD of hnRNPA1 increases the saturation concentration ($c_{sat}$) while addition of an hnRAC to the sequence reduces $c_{sat}$, suggesting a causative role for the motifs in LLPS [44].

The caveat of studies involving structural characterization of fibrils using solid-state methods is that they do not probe LCDs within the *liquid* environment of condensed droplets. If these fibrillar structures are in fact present within droplets, the protein fraction in droplets adopting these interactions is often low and, if maturation processes transpire, may grow over time [56]. Conversely, it has been argued that the fibrils offer an opportunity to trap pairwise interactions that are more transient in droplets. However, the interpretation of the fibril structure results are complicated by the fact that these β-sheet structures contain aromatic residues that have themselves been shown to drive LLPS [27,57] possibly without mediating extensive structuring. Additional work is needed to disentangle the importance of sequence features of LARKS and hnRACs *vs* the requirement of the structural motifs for function.

Whether the formation of fibrillar structures is required for LLPS or a consequence of the process thus remains an open question. Even if the fibrils emerge from liquid dense phases rather than mediate their formation, it is nonetheless crucial to characterize them structurally as they may provide insights into the conversion of liquid-like droplets into solid assemblies which have been implicated in numerous pathologies [37,58-61].

**Structures of low complexity domains studied in dense liquid phases**

A second set of reports has suggested that LCDs within condensed droplets remain entirely disordered and dynamic [62-65] (Figure 2). These conclusions stem from solution NMR spectroscopy on pure dense phases. In these studies, phase separation is induced in a large protein sample and all droplets are fused into one large drop by centrifugation or gravity.

The narrow chemical shift dispersion in the NMR spectra of these samples argue that the LCD in condensed liquid droplets remains disordered. No evidence for the formation of secondary structure is observed and relaxation experiments that probe the protein dynamics provide evidence for rapid local motions of the backbone and sidechains. Conversely, translational diffusion determined by pulse field gradient measurements is significantly slowed relative to that in the dilute phase [63,65], in line with observations from FRAP on individual microdroplets [38,62]. Comparably, NMR experiments performed on an elastin-like polypeptide showed that molecules within the dense phase exhibit disorder matching that observed in the dilute phase [66].

The slowed diffusion has been attributed to the high viscosity mediated by the high protein concentration within the dense phase (7-40 mM) [62-65] as well as the presence of weak multivalent interactions between the chains [63]. NOESY experiments that can selectively detect intermolecular interactions have detected close contacts between most residue types within the sequence of DDX4 and FUS [63,65]. Electrostatic and cation-$\pi$ interactions seem to drive phase separation of DDX4 while hydrophobic, $\pi$-$\pi$ and sp$^2$/$\pi$ interactions and hydrogen bonding likely occur in the condensed phase of FUS. The residue types mediating these interactions may therefore represent stickers in a disordered chain. Extensive NOE networks involving Gln residues were also observed in FUS; mutation of several di-Gln repeats across the sequence results in a reduced propensity for LLPS providing support for the driving force from these contacts for LLPS [65]. All-atom two-chain simulations suggested that hydrogen bonding between these residues may be important. Despite the lack of evidence for secondary structure, these hydrogen bonding patterns could be of a similar variety as those observed in the fibrillar structures, but they would have to be transiently sampled to result in agreement with the NMR results.

In addition to hydrogen boding, other interactions involving the same residues may be present simultaneously, such as sp$^2$/$\pi$ interactions between the sidechains of Gln residues, between Gln and Tyr residues and with the peptide backbone. Simulations also revealed the presence of hydrophobic and/or $\pi$-$\pi$ interactions between Tyr residues [65]. $\pi$-$\pi$ interactions between non-aromatic residues containing sp$^2$ groups such as Gln, Asn and the backbone may be important for phase separation of LCDs [67]. These conclusions were reached by the analysis of the frequencies of such interactions in ordered structures and the high fraction of $\pi$-containing residue types in phase-separating proteins. The dense phase of an elastin-like polypeptide shows evidence for the presence of hydrophobic interactions involving alanine, valine and proline based on intermolecular NOEs [66]. Many types of contacts have been observed, and the biophysical nature of the contacts dominating phase separation is expected to differ by sequence features of the respective LCD [68,69].

Intermolecular PRE experiments, which probe contacts over longer distances, provide further support that interactions are distributed across the sequence of FUS and hnRNPA2 such that all parts of the chain interact with all others, with little preference for contacts between particular regions. Given that PRE effects are present at distances of up to ~35 Å from the spin label, and the density of disordered protein chains within droplets may very well be high enough to approach one another within this range [63-65], these observations leave room for the driving force for phase separation stemming from stickers and spacers arranged along the disordered protein chain.

To date, the presence of ß-sheet structure within purely liquid droplets has not been shown. To test for the presence of ordered structure such as fibrils within the dense phase, Fawzi and coworkers performed NMR studies that probe dynamics across different timescales ranging from sub-microseconds to seconds [65]. None of the methods provided evidence for the presence of ordered structure, even at a low population. Considering the possibility that such a state might be invisible due to dynamics on a timescale that could not be probed by NMR, they turned to Raman spectroscopy but did not see evidence for the presence of fibrils. One would expect that if fibrillar structures were required for LLPS, they would be present at a sufficient abundance to be detected using these methods.

In summary, the currently available experimental data lend themselves to two opposing views with respect to the structural features of contacts within dense LCD phases. They either involve kinked cross-beta structure or interactions between largely disordered motifs. It has been argued that the interactions mediating fibrils could form transiently and exclusively pairwise to give rise to liquid droplets. While evidence argues for the importance of LARK/hnRAC sequence features for function, the same sequence features have also been implicated in the formation of critical contacts in disordered dense phases. Thus, whether LCD fibrils have physiological relevance for explaining liquid condensate structure and what the conformations of interacting motifs are within dense liquids remains to be fully determined.

While biomolecular condensates in cells behave as liquid-like bodies and are in that sense similar to simple in vitro droplets, condensates are complex, multicomponent systems. Different biomolecular condensates have different material properties, on the spectrum from liquid to solid, and these material properties may also evolve over time [5,6,70-72]. Not all proteins from which these properties emerge may sample the same structures and interactions, and they may not be fully captured within single-component systems. Nonetheless, characterizing the vast array of material states encompassing liquids, gels and solids is an important starting point to understanding how biomolecular condensates function in cells. Additionally, studies of LCDs within full-length proteins will be important. Multivalence via modular interaction domains and LCDs often coincide in phase-separating proteins, and the respective modulation of their properties needs to be characterized for a better understanding of the driving forces for phase separation.

**RNA structure in biomolecular condensates**

RNA is an important constituent in many biomolecular condensates and contributes to phase separation via RNA/RNA and protein/RNA interactions. RNA structure determination is challenging in itself [73], and thus determining the structures of RNA in dense phases and of the super-molecular structures formed presents even larger challenges. Hence, little data currently exists on RNA structure in biomolecular condensates. However, there is intriguing evidence that RNA structure can determine the identity of dense phases. The protein Whi3, a polyQ-rich RNA-binding protein from the filamentous fungus *Ashbya gossypii*, forms distinct RNA granules with different mRNAs. Careful analysis has revealed that the differences in RNA structures results in the immiscibility of the biomolecular condensates, and that denaturation of the RNAs results in a single type of condensate consisting of Whi3 and both types of RNAs [74]. While the structural characterization of the RNA was lacking atomistic detail and was not performed in the dense phase, the gain in knowledge regarding the molecular origin of specificity has been enormous, promising exciting future insights into the structure and function of RNA in dense phases.

**Structural biology meets phase separation – the future**

The evidence for the ubiquitous nature of LLPS throughout biology has been mounting quickly over the last few years, as has the realization that dysregulation of phase separation may be causal to diseases or accelerate their progression. These findings demand the attention of the biophysics and structural biology communities, as both can contribute to the quantitation of full phase behaviors, material properties and atomistic and super-molecular structures of dense phases. From the structural perspective, we should be striving to answer the following questions (Figure 3):

1. What are the atomistic details and the biophysical nature of the interactions within LCDs that drive phase separation?

2. Do the stabilities of domains and affinities of domain/motif interactions differ between dilute and dense phases? And are the structures of modular binding domain/motif interactions identical between dilute and dense phases, or are conformations affected by condensation?

3. To which extent do motifs/stickers in disordered regions become ordered upon interaction in the dense phase?

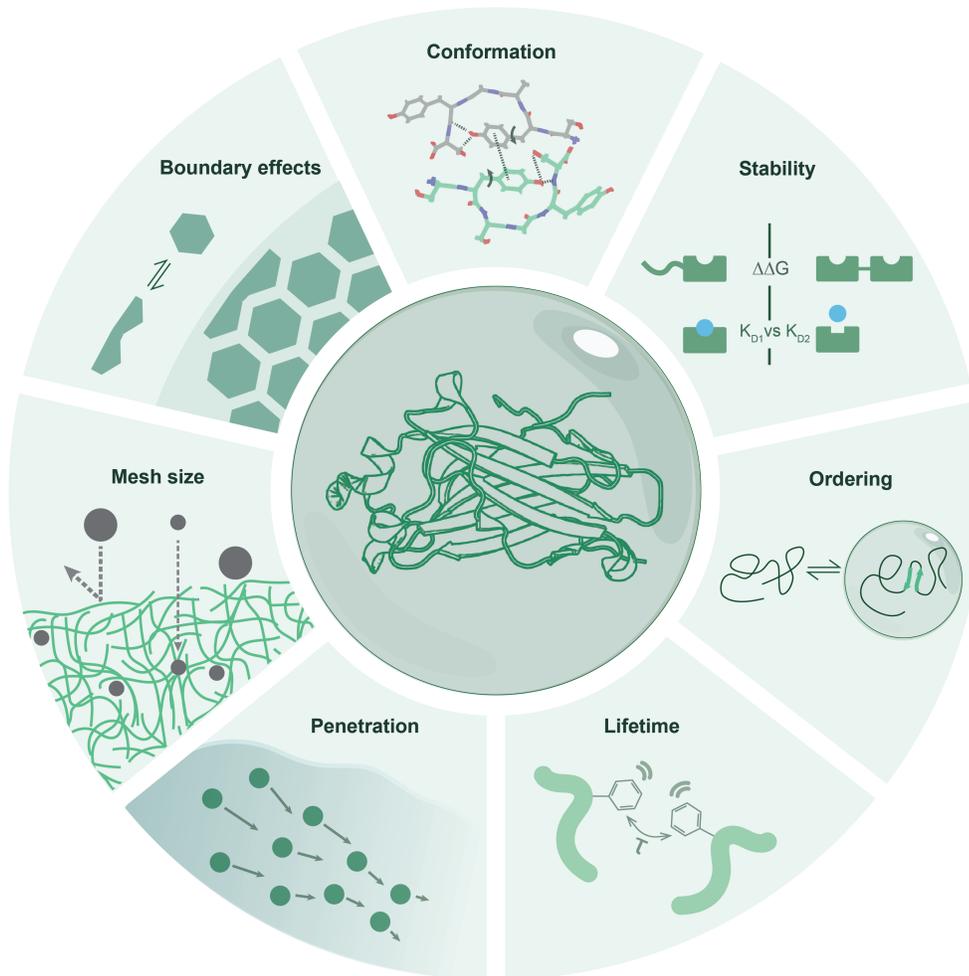

**Figure 3: (Super-)molecular structural properties that define dense phases.** Clockwise from the top: (1) What are the conformations adopted in phase separation-mediating contacts and what is the rearrangement of building blocks relative to each other (also see Figure 1)? (2) Do the stabilities of folded protein domains differ between the dilute and dense phase? And are the affinities between domains and motifs different between the phases? (3) What is the extent of ordering inside of droplets? (4) What are the lifetimes of individual interactions within the dense phase? (5) Over what distance within the dense phase is the movement of building blocks correlated? (6) How many cross-links form between molecules in the dense phase and what is the resulting mesh size? (7) Do the structures of molecules and their assemblies differ between the bulk dense phase and the phase boundary with the dilute phase?

4. What are the lifetimes of the interactions in dense phases?

5. How is the movement of molecules/building blocks in the dense phase correlated?

6. How many crosslinks do molecules form in the dense phase and what is the resulting mesh size?

7. Do the (super-)molecular structures differ between the bulk dense phase and the phase boundary with the dilute phase?

We should aim to tackle these questions in order to understand the molecular origin of material properties. They call for multi-pronged approaches that involve characterization of structure and dynamics on multiple length and time scales. The insight that LLPS plays fundamental roles in cell biology promises an understanding of cell biological processes from a solid biophysical, mechanistic basis. While the progress in our appreciation of the ubiquity of LLPS has been rapid, its characterization, with a few notable exceptions, has remained largely phenomenological. The opportunities for stringent quantitative biophysical and structural characterization are vast, and the potential gain in mechanistic understanding is as well. Such work will undoubtedly benefit from the ever-advancing tools of structural biology and keep the field engaged for years to come.

**Acknowledgements**
The authors thank Zhaowen Luo for help with Figure 3. T.M. acknowledges funding by NIH grant R01GM112846, St. Jude Children's Research Hospital, and the American Lebanese Syrian Associated Charities.